\documentclass[aps, prd, superscriptaddress, nofootinbib, preprintnumbers]{revtex4}
\usepackage{amsmath}
\usepackage{graphicx}

\newcommand{\lesssim}{\mathrel{\mathpalette\vereq<}}
\newcommand{\gtrsim}{\mathrel{\mathpalette\vereq>}}

\newcommand{\chushi}[1]{}

\begin{document}
 \preprint{MISC-2012-16}
 \title{{\bf Holographic techni-dilaton at 125 GeV}
 \vspace{5mm}}
\author{Shinya Matsuzaki}\thanks{
      {\tt synya@cc.kyoto-su.ac.jp} }
      \affiliation{ Maskawa Institute for Science and Culture, Kyoto Sangyo University, Motoyama, Kamigamo, Kita-Ku, Kyoto 603-8555, Japan.}
\author{{Koichi Yamawaki}} \thanks{
      {\tt yamawaki@kmi.nagoya-u.ac.jp}}
      \affiliation{ Kobayashi-Maskawa Institute for the Origin of Particles and 
the Universe (KMI) \\ 
 Nagoya University, Nagoya 464-8602, Japan.}
\date{\today}

\begin{abstract}
We find that a holographic walking technicolor model has a limit  (``conformal limit'') where 
the techni-dilaton (TD) becomes a 
massless Nambu-Goldstone boson 
of the scale symmetry with its nonzero finite
decay constant $F_\phi \ne 0$, which naturally realizes  a light TD, say at 125 GeV, near the limit.
In such a light TD case,  
we find that $F_\phi$ is uniquely determined by 
 the techni-pion decay constant 
$F_\pi$  independently of the holographic parameters: 
$F_\phi/F_\pi \simeq \sqrt{2 N_{\rm TF}}$, 
with $N_{\rm TF}$ being the number of techni-fermions. 
We show that the holographic TD is consistent with a new boson at 125 GeV 
recently discovered at the LHC. 
\end{abstract} 
\maketitle

\section{Introduction} 
A new boson of the mass around 125 GeV 
has recently been discovered at the LHC~\cite{ATLAS-CONF-2012,CMS-PAS-HIG}.  
It has been reported that in the diphoton channel 
the signal strength of the new boson is about two times larger than 
that predicted 
by the standard model (SM) Higgs, 
while other channels are consistent with the SM Higgs. 
This may imply a hint for a new scalar boson 
beyond the SM. For the theoretical possibilities, see,  for example, a recent review~\cite{Peskin:2012he}.

It is the techni-dilaton (TD) that 
is a candidate for such a new scalar boson: 
The TD is a composite scalar boson predicted in 
the walking technicolor (WTC)~\cite{Yamawaki:1985zg,Bando:1986bg}  
which is characterized by an approximately scale-invariant (conformal) 
 gauge dynamics and  a large anomalous dimension 
$\gamma_m =1$~\footnote{ 
The WTC was also studied subsequently without notion of anomalous dimension and  scale invariance/TD.~\cite{Akiba:1985rr}.}.    
The TD  arises as a pseudo Nambu-Goldstone (NG) boson for the spontaneous breaking of 
the approximate scale symmetry triggered by techni-fermion condensation. 
Its lightness, say 125 GeV, is therefore protected  
by the approximate scale symmetry inherent to the WTC.  
Thus the discovery of TD 
should imply discovery of the WTC.

In Refs.~\cite{Matsuzaki:2011ie,Matsuzaki:2012gd,Matsuzaki:2012vc,Matsuzaki:2012mk} 
the LHC signatures of the TD were studied.   
Particularly in Ref.~\cite{Matsuzaki:2012mk} (as well as Ref.~\cite{Matsuzaki:2012vc}) it was shown that the 125 GeV TD 
 is consistent with the currently reported diphoton signal as well as 
other signals such as $WW^*$ and $ZZ^*$, etc.. 
 It was emphasized that, in sharp contrast to other dilaton models~\cite{Goldberger:2007zk}
(See for example the recent analysis by Ref.~\cite{Ellis:2012hz}),
the TD is favored by the current data 
thanks to the presence of extra 
techni-fermion loop corrections to digluon and diphoton couplings.

The TD couplings to the SM particles take essentially the same form 
as those of  the SM Higgs. 
The overall scaling from the SM Higgs is just given by a ratio $v_{\rm EW}/F_\phi$, 
where $v_{\rm EW} (\simeq 246)$ GeV is the electroweak scale 
and $F_\phi$ denotes the TD decay constant which is in general $\neq v_{\rm EW}$~\footnote{ 
As was emphasized in Refs.~\cite{Matsuzaki:2011ie,Matsuzaki:2012gd,Matsuzaki:2012vc, Matsuzaki:2012mk}, 
the TD couplings to diphoton and digluon are not simply scaled from the SM Higgs, 
which include techni-fermion loop contributions depending on modeling of the WTC.}.  
The analysis of the previous works~\cite{Matsuzaki:2011ie,Matsuzaki:2012gd,Matsuzaki:2012vc,Matsuzaki:2012mk}  was  based 
on the evaluation of $F_\phi$ through the  assumption of the partially-conserved dilatation current (PCDC) which gives only a
combination $F_\phi^2  M_\phi^2$ in terms of the scale anomaly, where $M_\phi$ is the TD mass. The scale anomaly in turn was evaluated by the ladder
approximation, which was further related, through Pagels-Stokar formula for the techni-pion decay constant $F_\pi$,  to the electroweak scale 
$v_{\rm EW} = F_\pi \sqrt{N_D}$,
where $N_D$ is a number of weak doublet techni-fermions 
($N_D=4$ and $F_\pi \simeq 123$ GeV for the one-family model). 
Then we estimated up to the 30 \% uncertainties of the ladder 
approximations~\cite{Matsuzaki:2012mk} : 
 \begin{eqnarray} 
\frac{v_{\rm EW}}{F_\phi}
&\simeq& 
(0.1 - 0.3) \times  \left( \frac{N_D}{4} \right) \left( \frac{M_\phi}{125\,{\rm GeV}} \right) 
\,,\label{vals}
\end{eqnarray}
which was then shown to be consistent with the value of the best fit to 
the current LHC data in the case of the one-family model ($N_D=4$): 
 \begin{eqnarray} 
  \frac{v_{\rm EW}}{F_\phi}\Big|_{\rm best-fit} = 
\Bigg\{
\begin{array}{cc} 
0.22 & \qquad {\rm for} \qquad 
N_{\rm TC}=4 \\  
0.17  & \qquad {\rm for} \qquad N_{\rm TC}=5 
\end{array} 
\,. 
\label{best-fit}
  \end{eqnarray}

However, there is a potential problem in the ladder approximation about the mass of the TD as suggested earlier~\cite{Yamawaki:2007zz}: A straightforward calculation ~\cite{Harada:2003dc}
based on the  ladder Schwinger-Dyson (SD) equation and the ladder 
(homogeneous) Bethe-Salpeter (BS) equation in the walking regime indicates  a relatively light scalar bound state (identified with TD) as $M_\phi \sim 4 F_\pi$ ($\simeq 500$ GeV for the one-family model), which is much smaller than 
the techni-vector/axial-vector mesons on TeV range but still larger than the LHC boson at 125 GeV. This result~\cite{Harada:2003dc} is consistent with another calculation~\cite{Harada:2005ru} based on 
the ladder SD equation and the ladder (inhomogeneous) BS equation, and also consistent with other indirect computation \cite{Shuto:1989te} based on the ladder gauged Nambu-Jona-Lasinio model.
In fact the PCDC relation evaluated in the ladder approximation near the conformal window does not allow a very light  TD unless the TD gets decoupled with divergent decay constant~\cite{Haba:2010hu,Hashimoto:2010nw}:  The PCDC relation reads
\begin{eqnarray} 
  F_\phi^2 M_\phi^2    =  -4\langle \theta^\mu_\mu \rangle 
  = \frac{\beta(\alpha)}{\alpha} \langle G_{\mu \nu}^2 \rangle
 \simeq  3 \eta  m_F^4, 
  \label{PCDC:1}
   \end{eqnarray}
where $\langle G_{\mu\nu}^2 \rangle$ is the techni-gluon condensate 
with $\beta(\alpha)$ being a beta function of the TC gauge coupling $\alpha$ and 
the last equation is the ladder estimate near the conformal window with 
$\eta \simeq  \frac{N_{\rm TC} N_{\rm TF}}{2\pi^2}  ={\cal O} (1)$~\cite{Hashimoto:2010nw,Miransky:1989qc} 
(For earlier references, see \cite{Bardeen:1985sm}). 
This simply implies $(F_\phi/m_F)^2 \cdot (M_\phi/m_F)^2 \rightarrow$ constant $\ne 0$ near the conformal window $m_F/\Lambda \rightarrow 0$, with $\Lambda$ being the analogue of the $\Lambda_{\rm QCD}$,  
the intrinsic scale of the walking technicolor where the infrared conformality terminates beyond that scale.
Then the limit $M_\phi/m_F \rightarrow 0 $,  where  the TD gets light compared with the weak scale $m_F  (={\cal O} (4 \pi F_\pi) )$,  can only be realized when $F_\phi/m_F  \rightarrow \infty$, i.e., a decoupled limit.

A possible way out would be to include fully {\it nonperturbative gluonic dynamics}. Actually,
the ladder approximation totally ignores non-ladder dynamics most notably the full gluonic dynamics.
Also a direct estimate of $F_\phi$ free from the ladder approximation and 
without invoking the PCDC (without referring to $M_\phi$)  
is necessary to give more implications of the TD at the LHC.  
One such a possibility besides lattice simulations would be a holographic computation based on 
the gauge-gravity duality~\cite{Maldacena:1997re}.

In this paper, we make a full analysis of a holographic model dual to the WTC previously 
proposed in Ref.~\cite{Haba:2010hu} 
by including the bulk field dual to the techni-glueball so as to incorporate 
the fully nonperturbative gluonic dynamics. 
We show that thanks to the nonperturbative gluonic dynamics in contrast to the ladder approximation, we do have an  exactly massless TD limit (``conformal limit")  
: 
\begin{equation} 
\frac{M_\phi}{F_\pi} \to 0 
\qquad 
{\rm with} 
\qquad 
\frac{F_\phi}{F_\pi} = {\rm finite}
\,.  
\end{equation}
The resultant $F_\phi$ is fairly independent of the TD mass $M_\phi$, 
in contrast to the PCDC estimation in the ladder approximation.  
Remarkably enough, in the light TD case, 
we find a novel relation between $F_\phi$ and the techni-pion decay constant $F_\pi$, 
independently of the holographic parameters: 
\begin{equation} 
\frac{F_\phi}{F_\pi}  
\simeq \sqrt{2 N_{\rm TF}}
\,, 
\end{equation} 
with $N_{\rm TF}$ being the number of techni-fermions. 
 In such a light TD limit the masses of techni-$\rho$ ($M_\rho$) and -$a_1$ $(M_{a_1})$ mesons 
also go to zero,  $M_{\rho,a_1}/F_\pi \to 0$, 
which implies a scaling property similar to 
the vector realization~\cite{Georgi:1989gp} and the vector manifestation~\cite{Harada:2000kb} based on 
the hidden local symmetry~\cite{Harada:2003jx}.

We discuss the 125 GeV holographic TD at the LHC taking the one-family model as a definite benchmark. 
The TD couplings to the SM particles set by the ratio $v_{\rm EW}/F_\phi$ are 
estimated, say, for $N_{\rm TC}=4$ and $N_{\rm TF} = 16, 20$,  
to be $v_{\rm EW}/F_\phi \simeq 0.2$ (up to $1/N_{\rm TC}$ corrections), 
which turns out to be on the best-fit value in Eq.(\ref{best-fit}) favored by the current data on 
a new boson at 125 GeV recently observed at the LHC~\cite{ATLAS-CONF-2012,CMS-PAS-HIG}
(See Table~\ref{tab:chi2}).

This paper is organized as follows: 
In Sec.~\ref{sec:model} 
we start with a brief review of the holographic WTC model proposed in Ref.~\cite{Haba:2010hu} 
to explain the holographic computation of the chiral and gluon condensates (Sec.~\ref{sec:condensates}), 
current correlators and masses of the related lightest resonances, 
$M_\rho$, $M_{a_1}$, $M_\phi$ and techni-glueball $M_G$ in the WTC 
(Secs.~\ref{sec:PiVPiA} and \ref{sec:PiSPiG}).    
In Sec.~\ref{sec:HTD125} 
 we next turn to computation of the TD decay constant $F_\phi$, which can actually be done 
by combining the Ward-Takahashi identities for the dilatation and scalar currents (Sec.~\ref{sec:Fphi}). 
We then discuss the light TD case and show that the 
massless Nambu-Goldstone boson 
limit (``conformal limit")  can be realized in the present model. 
In such a light TD case, we find a novel relation between $F_\phi$ and $F_\pi$, which 
is independent of the holographic-model parameters, to be just a constant (Sec.~\ref{sec:LTDlimit}).      
In Sec.~\ref{sec:HTD125} we discuss the 125 GeV holographic TD at the LHC and 
show that the TD is consistent with a new boson at 125 GeV currently reported from the LHC experiments. 
Sec.~\ref{sec:summary} devotes to summary of this paper.

\section{Model}
\label{sec:model}

The holographic model proposed in Ref.~\cite{Haba:2010hu} 
is based on deformation of 
a bottom-up approach for successful holographic-dual of QCD~\cite{DaRold:2005zs,Erlich:2005qh} 
 with $\gamma_m\simeq 0$, which is extended to WTC~\cite{Hong:2006si,Piai:2006hy,Haba:2008nz} with $\gamma_m \simeq 1$. 
The model describes a five-dimensional gauge theory having 
$SU(N_{\rm TF})_L \times SU(N_{\rm TF})_R$ gauge symmetry, 
defined on the five-dimensional 
anti-de-Sitter space (AdS$_5$) with $L$, the curvature radius of AdS$_5$, 
described by the metric $ds^2= g_{MN} dx^M dx^N 
= \left(L/z \right)^2\big(\eta_{\mu\nu}dx^\mu dx^\nu-dz^2\big)$ 
with $\eta_{\mu\nu}={\rm diag}[1, -1, -1,-1]$. 
The fifth direction $z$ is compactified on an interval extended 
from the ultraviolet (UV) brane located at $z=\epsilon$ 
to the infrared (IR) brane at $z=z_m$, i.e., $ \epsilon \leq z \leq z_m  $.  
  In addition to the bulk left- ($L_M$) and right- ($R_M$) gauge fields,  
we introduce a bulk scalar field $\Phi_S$ which transforms as bifundamental representation under 
the $SU(N_{\rm TF})_L \times SU(N_{\rm TF})_R$ gauge symmetry 
so as to deduce the information concerning 
the techni-fermion bilinear operator $\bar{F} F$. 
The mass-parameter $m_{\Phi_S}$ is then related to $\gamma_m$ 
as $m_{\Phi_S}^2=- (3-\gamma_m)(1+ \gamma_m)/L^2$, 
where $\gamma_m \simeq 1$.  
An extra bulk scalar field $\Phi_G$ 
dual to techni-gluon condensate $\langle \alpha G_{\mu\nu}^2
\rangle$ is incorporated, 
where $\alpha$ is related to the TC gauge couping $g_{\rm TC}$ by $\alpha = g_{\rm TC}^2/(4\pi)$. 
Because $\langle \alpha G_{\mu\nu}^2 \rangle$ is singlet under 
the chiral $SU(N_{\rm TF})_L \times SU(N_{\rm TF})_R$ symmetry, 
the dual-bulk scalar $\Phi_G$ has to be a real field. 
We take ${\rm dim}(\alpha G_{\mu\nu}^2)=4$ and 
the corresponding bulk-mass parameter $m_{\Phi_G}^2=0$.

 The action in Ref.~\cite{Haba:2010hu} is thus given as  
\begin{equation} 
  S_5 = S_{\rm bulk} + S_{\rm UV} + S_{\rm IR} 
  \,, \label{S5}
\end{equation}
where $S_{\rm bulk}$ denotes the five-dimensional bulk action, 
\begin{eqnarray} 
  S_{\rm bulk} 
  &=& 
  \int d^4 x \int_\epsilon^{z_m} dz 
  \sqrt{-g} 
  \frac{1}{g_5^2} \, e^{c_G g_5^2 \Phi_G} 
 \Bigg[ 
\frac{1}{2} \partial_M \Phi_G \partial^M \Phi_G 
\nonumber \\ 
&& 
+ {\rm Tr}[D_M \Phi_S^\dag D^M \Phi_S - m_{\Phi_S}^2 \Phi_S^\dag \Phi_S ] 
\nonumber \\ 
&&
  - \frac{1}{4} {\rm Tr}[L_{MN}L^{MN} + R_{MN}R^{MN}] 
 \Bigg] 
 \,, \label{S:bulk}
\end{eqnarray}
and $S_{\rm UV, IR}$ the boundary actions,  
\begin{eqnarray} 
S_{\rm UV} 
&=&  
   \int d^4 x \int_\epsilon^{z_m} dz  \, \delta (z -\epsilon) 
  \sqrt{-\tilde{g}} \, {\cal L}_{\rm UV} 
\,, \nonumber \\ 
 S_{\rm IR} 
&=&  
   \int d^4 x \int_\epsilon^{z_m} dz  \, \delta (z -z_m) 
  \sqrt{-\tilde{g}} \, {\cal L}_{\rm IR} 
\,, \label{S:UVIR}
\end{eqnarray}
with the boundary-induced metric $\tilde{g}_{\mu\nu}=(L/z)^2 \eta_{\mu\nu}$. 
In Eq.(\ref{S:bulk}),  
the covariant derivative acting on $\Phi_S$ is defined as 
$D_M\Phi_S=\partial_M \Phi_S+iL_M\Phi_S-i\Phi_S R_M$;   
$L_M(R_M)=L_M^a(R_M^a) T^a$ with the generators of $SU(N_{\rm TF})$ normalized by 
${\rm Tr}[T^a T^b]=\delta^{ab}$; 
$L(R)_{MN} = \partial_M L(R)_N - \partial_N L(R)_M 
 - i [ L(R)_M, L(R)_N ]$; 
$g={\rm det}[g_{MN}]=-(L/z)^{10}$; 
the gauge coupling $g_5$ and a parameter $c_G$ 
are fixed by the desired UV asymptotic forms of the vector/axial-vector current correlator 
to be~\cite{Haba:2010hu} 
\begin{eqnarray} 
\frac{L}{g_5^2} &=& 
\frac{N_{\rm TC}}{12\pi^2}  
\,, \nonumber \\ 
c_G &=& 
- \frac{N_{\rm TC}}{192 \pi^3} 
\,. 
\end{eqnarray}

The UV boundary action $S_{\rm UV}$ in Eq.(\ref{S:UVIR}) plays a role of 
the UV regulator to absorb the UV-divergent $\epsilon$ terms 
arising from the five-dimensional bulk dynamics, which we will not specify. 
The IR boundary action $S_{\rm IR}$ is introduced so as to 
realize minimization of the bulk potential by 
nonzero chiral condensate~\cite{Haba:2010hu} with the IR Lagrangian: 
\begin{eqnarray} 
  {\cal L}_{\rm IR} 
  &=& - \chi^2 \left( m_b^2 {\rm Tr}[|\Phi_S|^2] + \lambda {\rm Tr}[|\Phi_S|^2]^2 
\right)
\,, \nonumber \\ 
{\rm with} \qquad  
\chi 
&=&e^{\frac{c_G g_5^2}{2} \Phi_G}
\,. \label{L:IR}
\end{eqnarray}

\subsection{Condensates} 
\label{sec:condensates}

The bulk scalar fields $\Phi_S$ and $\Phi_G$ (or $\chi$) are parametrized as follows: 
\begin{eqnarray} 
\Phi_S(x,z) 
&=& \frac{1}{\sqrt{2}} \left[ v(z)+ \frac{\sigma(x,z)}{\sqrt{N_{\rm TF}}} \right] e^{i \pi(x,z)/v(z)}
\,, \nonumber \\ 
\chi(x,z) &=& 
v_{\chi}(z) e^{\sigma_\chi(x,z)/v_{\chi}(z)}    
\,,  
\end{eqnarray} 
with the vacuum expectation values (VEVs), $v(z)=\sqrt{2} \langle \Phi_S \rangle$ and 
 $v_{\chi}(z)= \langle \chi \rangle$, respectively. 
Since the techni-pions tend to be on the order of several hundred GeV~\cite{Jia:2012at} and hence do  
not directly affect the TD phenomenology at the LHC,  
in the present study we will disregard techni-pions $\pi(x,z)$.  
The boundary condition for $v(z)$ is chosen~\cite{Haba:2008nz}:  
\begin{eqnarray}  
v(\epsilon)   
&=& \left(\frac{\epsilon}{L} \right)^2 \log \frac{z_m^2}{\epsilon^2} \,  c_S M
\,, \nonumber \\   
v(z_m) &=& \frac{\xi}{L}
\,, \label{BC:v}
\end{eqnarray}  
where $M$ stands for the current mass of techni-fermions and 
the IR value $\xi$ is related to the techni-fermion condensate $\langle \bar{F}F \rangle_{1/L}$ 
renormalized at the scale $\mu=1/L$~\cite{Haba:2008nz}. 
The intrinsic log factor in the UV boundary condition Eq.(\ref{BC:v}) has been 
supplied in order to smoothly connect the chiral condensate at $\gamma_m =1$ to that for  $\gamma_m \lesssim 1$~\cite{Haba:2008nz}. 
The parameter $c_S$ has been introduced which can arise 
from the ambiguity of the definition for the current mass $M$ and 
is to be fixed to be $c_S = \sqrt{3}/2$ for $\gamma_m \simeq 1$, 
by matching the UV asymptotic form of the scalar current correlator 
to the form predicted from the operator product expansion,  
as will be clarified later (See Eq.(\ref{cS})).

The boundary condition for $v_\chi$ is taken as    
\begin{eqnarray} 
\lim_{\epsilon \to 0} v_\chi(z) |_{z=\epsilon}
&=& 
e^{\frac{c_G}{2} \frac{g_5^2}{L} M'} = e^{-\frac{1}{32\pi} L M'}
\,,  \nonumber \\  
v_\chi(z)|_{z=z_m} &=& 1 + G
\,, 
\end{eqnarray}  
where $M'$ becomes the external source for the techni-gluon 
condensation-operator $(\alpha G^2_{\mu\nu})$ 
and $G$ is associated with the techni-gluon condensate 
$ \langle \alpha  G_{\mu\nu}^2 \rangle$ 
($G \simeq 0.25$ in the case of the real-life QCD)~\cite{Haba:2010hu}.

Solving the equations of motion for these VEVs and 
putting their solutions back into the action $S_5$ in Eq.(\ref{S5}), 
one can calculate the chiral and 
gluon condensates ($\langle \bar{F}F \rangle$  and $\langle \alpha G^2_{\mu\nu} \rangle$) 
based on the holographic recipe (For details, see Ref.\cite{Haba:2010hu})~\footnote{
The nonzero chiral condensate $(\xi)$ can be ensured thanks to the presence of 
the IR boundary potential in Eq.(\ref{L:IR}), such that 
$\xi$ is related to other IR values in Eq.(\ref{L:IR}) as 
$\xi^2 = 1/\lambda \left[ (m_b L)^2  - N_{\rm TC}/(6\pi^2) (1-G)/(1+G) \right]$~\cite{Haba:2010hu}. 
}:  
\begin{eqnarray} 
- \lim_{\epsilon \to 0} \frac{\delta S_5}{\delta M} \Bigg|_{M=0}
&=& \langle \bar{F} F \rangle_{1/L} 
\,, \nonumber \\ 
&=& 
 -\frac{c_{S} N_{\rm TF} N_{\rm TC}}{6 \pi^2} \frac{\xi (1+G)}{z_m^3} \left( \frac{L}{z_m} \right)^{-1}
\,, \nonumber \\ 
- \lim_{\epsilon \to 0} \frac{\delta S_5}{\delta M'} \Bigg|_{M'=0} 
&=& \langle \alpha G_{\mu\nu}^2 \rangle
\nonumber \\ 
&=&  
\frac{32 N_{\rm TC}}{3\pi} \frac{G}{z_m^4}
\,. \label{FFbar}
\end{eqnarray}

\subsection{Current correlators}
\label{sec:Pis}

We calculate current correlators in the scalar sector as well as the vector and axial-vector sectors   
by extending the analysis in Ref.~\cite{Haba:2010hu}. 
  For that purpose, 
it is necessary to specify the boundary conditions for 
$\sigma, \sigma_\chi, L_M$ and $R_M$.  
We first consider the UV boundary condition for $\sigma$, which 
is assigned similarly to $v(z)$ in Eq.(\ref{BC:v}), 
\begin{equation} 
\sigma(x, \epsilon)   
= \left( \frac{\epsilon}{L} \right)^2 \log \frac{z_m^2}{\epsilon^2} \, c_S s(x) 
\, ,
\end{equation} 
with $s(x)$ being a source for the scalar current $J_S=\bar{F}F/\sqrt{N_{\rm TF}}$. 
The IR boundary condition is chosen in such a way that 
the terms in quadratic order of $\sigma$ vanish at the IR boundary including the 
IR boundary potential Eq.(\ref{L:IR})~\cite{Haba:2010hu},     
\begin{equation} 
\left[ 
\partial_z +  2g_5^2 \frac{L}{z} \left( 
\lambda \frac{\xi^2}{L^2} - \frac{1}{g_5^2 L} \frac{1-G}{1+G}  \right) 
\right] 
\sigma(x, z) \Bigg|_{z=z_m} = 0 
\,. 
\end{equation}  
Similarly, one can impose the boundary condition for $\sigma_\chi$. 
It turns out that the boundary condition should be 
\begin{eqnarray} 
\sigma_\chi(x,\epsilon) 
&=& \frac{g(x)}{L}
\,, \nonumber \\   
\partial_z \sigma_\chi(x,z)|_{z=z_m} 
&=& 0
\,,  
\end{eqnarray} 
where $g(x)$ denotes 
the source for the current correlator for the techni-gluon condensation operator,  
$J_G= \alpha G_{\mu\nu}^2$.

Next, consider the vector and axial-vector sectors. 
One defines the five-dimensional vector and axial-vector gauge fields 
$V_M$ and $A_M$ as  
$V_M=(L_M+R_M)/\sqrt{2} $ and  $A_M= (L_M-R_M)/\sqrt{2}$. 
The UV boundary values of $V_\mu$ and $A_\mu$ then 
play the role of the sources ($v_\mu$, $a_\mu$) 
for the vector $(J_V^{\mu a}=\bar{F} \gamma^\mu T^a F)$ and 
axial-vector $(J_A^{\mu a}=\bar{F} \gamma^\mu  \gamma_5 T^a F)$  currents 
externally coupled to the WTC sector. 
By working in $V_z = A_z \equiv 0$ gauge, 
their boundary conditions are chosen as  
$
\partial_z V_{\mu}(x,z) |_{z=z_m} =
\partial_z A_{\mu}(x,z) |_{z=z_m}=0 
$,  
$
V_{\mu}(x,z) |_{z=\epsilon}=v_{\mu}(x) 
$, and 
$
A_{\mu}(x,z)\big|_{z=\epsilon}=a_{\mu}(x) 
$.

One can thus calculate the scalar, gluon, vector and axial-vector current 
correlators $\Pi_S, \Pi_G, \Pi_V$ and $\Pi_A$, respectively, as follows: 
\begin{eqnarray} 
-  \frac{\delta^2 S_5}{\delta s(-q) \delta s(q)} \Bigg|_{s=0}
&=& 
i\int d^4 x \,e^{iq\cdot x} \langle 
J_S (x) J_S(0) \rangle 
\nonumber \\ 
&=&
\Pi_S(-q^2) 
\,, \nonumber \\ 
-  \frac{\delta^2 S_5}{\delta g(-q) \delta g(q)} \Bigg|_{g=0}
&=& 
i\int d^4 x \,e^{iq\cdot x} \langle 
J_G (x) J_G(0) \rangle 
\nonumber \\ 
&=& 
\Pi_G(-q^2) 
\,, \nonumber \\ 
\frac{ \delta^2 S_5}{\delta v^a_\mu(q) \delta  v^b_\nu(-q)} \Bigg|_{v=0} 
&=& 
i\int d^4 x \,e^{iq\cdot x} \langle 
J^{a\mu}_V (x) J^{b\nu}_V(0) \rangle 
\nonumber \\ 
&=& 
-\delta^{ab} \left(\eta^{\mu\nu}-\frac{q^{\mu}q^{\nu}}{q^2}\right)\,  \Pi_{V}(-q^2) 
\, , 
\nonumber \\ 
\frac{ \delta^2 S_5}{\delta a^a_\mu(q)  \delta a^b_\nu(-q)} \Bigg|_{a=0} 
&=& 
i\int d^4 x \,e^{iq\cdot x} \langle 
J^{a\mu}_A (x) J^{b\nu}_A(0) \rangle 
\nonumber \\ 
&=& 
-\delta^{ab} \left(\eta^{\mu\nu}-\frac{q^{\mu}q^{\nu}}{q^2}\right)\, \Pi_{A}(-q^2) 
\, .  
\nonumber \\ \label{correlators}
\end{eqnarray}

\subsection{$\Pi_V$ and $\Pi_A$} 
\label{sec:PiVPiA}

The vector and axial-vector current correlators $\Pi_V$ and $\Pi_A$ 
can be expanded in terms of towers of 
the vector and axial-vector resonances with the masses $M_{V_n,A_n}$ 
and decay constants $F_{V_n,A_n}$ as 
\begin{equation} 
  \Pi_{V,A}(q^2) =  
\sum_n \frac{F_{V_n,A_n}^2 M_{V_n,A_n}^2}{M_{V_n,A_n}^2 - q^2}
  \,. 
\end{equation} 
We identify the lowest poles for $\Pi_{V,A}$ as the techni-$\rho$ and -$a_1$ mesons. 
Their masses $M_{V_1} \equiv M_\rho$ and $M_{A_1} \equiv M_{a_1}$ 
 are calculated through solving the eigenvalue equations for 
the vector and axial-vector profile functions 
 $V_1(z)$ and $A_1(z)$~\cite{Haba:2010hu}:  
\begin{eqnarray} 
&& 
 \left[ 
 M_\rho^2 + \omega^{-1}(z) \partial_z \omega(z) \partial_z  
\right] V_1(z) = 0 
\,, \nonumber \\ 
&& 
\Bigg[ 
  M_{a_1}^2 + \omega^{-1}(z) \partial_z \omega(z) \partial_z  
- 2 \left( \frac{L}{z} \right)^2 v^2(z) 
\Bigg] A_1(z)=0
\,, \label{profile:eq}
\end{eqnarray}
with $\omega(z) = (L/z) v_\chi^2(z)$ and 
the boundary condition $V_1(\epsilon)=0$, $\partial_z V_1(z_m)=0$ and similar one for $A_1(z)$.  
 Using the solutions of the VEVs in the limit where $M \to 0$ and $M' \to 0$,  
\begin{eqnarray}  
v_\chi(z) &=& 
1 + G \left( \frac{z}{z_m} \right)^4 
\,, \nonumber \\ 
v(z) &=& 
\frac{\xi}{L} \frac{1+G}{1+G(z/z_m)^4} 
\frac{\log(z/\epsilon)}{\log(z_m/\epsilon)} 
\end{eqnarray} 
we find 
$M_\rho$ and $M_{a_1}$ as a function of just two parameters $\xi$ and $G$ 
with the overall scale set by $z_m$: 
\begin{eqnarray} 
  M_\rho &=& 
z_m^{-1} \cdot \widetilde{M}_\rho(G) 
  \,, \nonumber \\ 
  M_{a_1} &=& 
  z_m^{-1} \cdot \widetilde{M}_{a_1} (\xi, G) 
  \,. \label{Mrho:Ma1}
\end{eqnarray}

In addition,  from $\Pi_V$ and $\Pi_A$ we may construct 
the $S$ parameter: 
\begin{equation}
S=-4\pi \, N_D \, \frac{d}{ d
Q^2}\left[\Pi_V(Q^2)-\Pi_A(Q^2)\right]_{Q^2=0} 
\,, 
\end{equation}
where $Q \equiv \sqrt {- q^2}$ and 
$N_D$ denotes the number of electroweak doublets. 
 Once $N_{\rm TC}$ and $N_D$ are given, 
the present holographic model allows us to calculate $S$ as a function of just two parameters  
 $\xi$ and $G$~\cite{Haba:2010hu}:  
\begin{eqnarray} 
 S &=& \frac{N_D N_{\rm TC}}{3\pi} \int_{t_\epsilon}^1 \frac{dt}{t} v_\chi^2(t) \left[ 1 - A^2(t) \right]  
   \nonumber \\ 
   &\equiv & N_D \cdot \widehat{S}(\xi, G; N_{\rm TC})
 \,, \label{Spara}
\end{eqnarray}
 where $t_\epsilon = \epsilon/z_m (\to 0)$ and $A(t)$ satisfies the second equation in 
Eq.(\ref{profile:eq}) with the zero momentum $q=M_{a_1}=0$ set.

We also introduce the techni-pion decay constant defined as 
\begin{equation} 
  F_\pi^2 = \Pi_V(0) - \Pi_A(0)  
  \,, \label{fpi:def}
\end{equation}
which is related to the electroweak scale $v_{\rm EW}$ as $F_\pi = v_{\rm EW}/\sqrt{N_D}$. 
  The present model enables us to calculate $F_\pi$ as a function of $\xi$, $G$ and $z_m$  
for given $N_{\rm TC}$~\cite{Haba:2010hu}: 
\begin{eqnarray} 
  F_\pi^2 &=&  
 \frac{N_{\rm TC}}{12\pi^2} \frac{\widetilde{F}^2(\xi, G)}{z_m^2} 
 \,, \label{Fpi}
\end{eqnarray} 
where 
$\widetilde{F}^2 = \partial_t A(0,t)/t|_{t=t_\epsilon \to 0} $.

\subsection{$\Pi_S$ and $\Pi_G$} 
\label{sec:PiSPiG}

The scalar current correlator $\Pi_S$ is straightforwardly evaluated through Eq.(\ref{correlators}). 
 In calculating $\Pi_S$ we encounter some divergent terms arising by taking $\epsilon \to 0$, 
which can be renormalized by the UV boundary action in Eq.(\ref{S:UVIR}). 
Letting such a ``bare" correlator be $\Pi_S|_{1/\epsilon}$ and 
renormalizing it at $\mu=1/L$ as $\Pi_S|_{1/\epsilon} = (\epsilon/L) \Pi_S|_{1/L}$, 
we arrive at  
\begin{equation} 
  \Pi_S(q^2)|_{1/L} = - c_S^2 \cdot \frac{N_{\rm TC}}{6 \pi^2}\left( \frac{1}{qL}  \right)^2 q^2 
\left[ \log (qL)^2 - \pi \Xi (q) \right]   
\,, \label{PiS}
\end{equation}  
where 
\begin{eqnarray} 
  \Xi(q) &=&
 \frac{A \cdot Y_0(qz_m) - qz_m Y_1(qz_m)}{A \cdot J_0(qz_m) - qz_m J_1(qz_m)}
\,, \nonumber \\ 
A &=& 
\frac{24 \pi^2 \lambda \xi^2}{N_{\rm TC}} 
= \frac{3}{2} \kappa \xi^2 
\,, 
\end{eqnarray}
with $J_{0,1}$ and $Y_{0,1}$ being the Bessel functions. 
 Here we have used $\lambda = \kappa N_{\rm TC}/(4\pi)^2$ where we set $\kappa =1$~\cite{Haba:2010hu}. 
 The UV asymptotic form of Eq.(\ref{PiS}) may be compared with the operator-product expansion form: 
\begin{equation} 
  \Pi_S(q^2)|_{1/L} = \left( \frac{1}{qL}  \right)^2 q^2 \left[ - \frac{N_{\rm TC}}{8\pi^2} q^2 \log(q L)^2  
+ \cdots \right]
\,, 
\end{equation}   
such that we find the matching condition for the model parameter $c_S$~\footnote{
In the previous analysis~\cite{Haba:2010hu}, 
without explicit evaluation of $\Pi_S$ in the case of WTC with $\gamma_m \simeq 1$,   
the parameter $c_S$ was set to $\sqrt{3}$ simply taken from the QCD case with $\gamma_m \simeq 0$~\cite{DaRold:2005vr}. 
}, 
\begin{equation} 
  c_S = \frac{\sqrt{3}}{2} 
 \,. \label{cS}
\end{equation}

The scalar current correlator $\Pi_S$ can also be expressed 
in terms of tower of the scalar resonances with the masses $M_{S_n}$ and decay constants 
$F_{S_n}$: 
\begin{equation}
    \Pi_S(q^2) = \sum_n \frac{F_{S_n}^2 M_{S_n}^2}{M_{S_n}^2 - q^2} 
    \,. \label{PiS:gene}
\end{equation}
Using this and Eq.(\ref{PiS}) we extract the scalar masses and the scalar decay constants renormalized 
at $\mu=1/L$ as 
\begin{eqnarray} 
   M_{S_n} &:&  
\frac{3}{2}\kappa \xi^2 \cdot J_0(M_{s_n} z_m) = M_{S_n} z_m J_1(M_{S_n} z_m) 
\,, \nonumber \\ 
   F_{S_n}^2|_{1/L}  &=& 
 \frac{N_{\rm TC}}{2\pi^2} \frac{1}{z_m^2} \left( \frac{1}{M_{S_n} L} \right)^2  
   \frac{1}{J_0^2(M_{S_n} z_m) + J_1^2(M_{S_n} z_m)} 
\,. \label{MS:FS}
\end{eqnarray}

Similarly, we can calculate the current correlator for the gluon-condensation operator $\Pi_G$ 
to find the masses and decay constants associated with the resonances arising in $\Pi_G$: 
\begin{eqnarray} 
 M_{G_n} &=& \frac{j_{1,n}}{z_m} 
\,, \nonumber \\ 
 F_{G_n}^2 &=&  \frac{128 N_{\rm TC}}{3}  \frac{1}{z_m^2} \frac{1}{J^2_0(M_{G_n} z_m)}
\,, \label{MG:FG}  
\end{eqnarray}
where $j_{1,n}$ denotes the $n$th zero of the Bessel function $J_1$. 
We identify the lowest resonance in $\Pi_G$ as the techni-glueball ($G$), 
i.e, $M_{G_1} \equiv M_G$ and $F_{G_1} \equiv F_G$.

\subsection{Techni-dilaton decay constant} 
\label{sec:Fphi}

 We next compute the TD decay constant $F_\phi$ from the present holographic model. 
To this end,  following Ref.~\cite{Hashimoto:2011ma} we start with the Ward-Takahashi identity 
for the dilatation current $D_\mu$ coupled to techni-fermion bilinear operator $\bar{F}F$: 
\begin{eqnarray} 
&&\lim_{q_\mu \to 0} \int d^4 x e^{iqx} \langle 0| T \theta_\mu^\mu(x) \bar{F}F(0) | 0 \rangle 
\nonumber \\ 
&& 
=
- (3-\gamma_m) \langle 0| \bar{F}F |0 \rangle 
\,, \label{WT}
\end{eqnarray}
where $(3-\gamma_m) \simeq 2$ and $\theta_\mu^\mu = \partial_\mu D^\mu$.  
The TD arises as the lightest scalar which couples to 
the dilatation current $D_\mu$ with the coupling strength $F_\phi$ at the on-shell $p^2=M_\phi^2$: 
\begin{equation} 
  \langle 0 | \theta_\mu^\mu(0) | \phi \rangle = F_\phi M_\phi^2 
\,.   
\end{equation}
  The TD pole therefore contributes to the left-hand side of Eq.(\ref{WT}) such that 
\begin{equation} 
  F_\phi \langle \phi(q=0) | \bar{F}F(0) |0\rangle 
=  - (3-\gamma_m) \langle \bar{F}F \rangle 
\,. \label{TD:WT}
\end{equation}
Since the TD couples also to the scalar current $J_S=\bar{F}F/\sqrt{N_{\rm TF}}$, 
we may define the amplitude: 
\begin{equation} 
  \langle \phi(q=0) | J_S(0) |0\rangle = F_S M_\phi 
\,. \label{TD:JS}
\end{equation}
   Comparing this with the spectral representation of $\Pi_S$ in Eq.(\ref{PiS:gene}), 
we may identify the lightest scalar arising in $\Pi_S$ as the TD, i.e., $M_{S_1}\equiv M_\phi$ 
and $F_{S_1} \equiv F_S$.  
  From Eqs.(\ref{TD:WT}) and (\ref{TD:JS}), we thus construct the TD decay constant $F_\phi$ as 
\begin{equation} 
  F_\phi = \frac{- 2 \langle \bar{F}F  \rangle}{\sqrt{N_{\rm TF}} F_{S} M_\phi}
  \,. \label{Fphi:formula}
\end{equation} 
 Note that this $F_\phi$ is renormalization-scale independent as it should be: 
 $\frac{\langle \bar{F}F \rangle_{1/L}}{F_S|_{1/L}} = \frac{\langle \bar{F}F \rangle_{M_\phi}}{F_S|_{M_\phi}}$. 
Putting Eqs.(\ref{FFbar}) and (\ref{MS:FS}) into Eq.(\ref{Fphi:formula}) 
we now obtain the holographic formula for $F_\phi$, 
\begin{equation} 
 F_\phi 
= \sqrt{\frac{N_{\rm TF} N_{\rm TC}}{6\pi^2} [ J_0^2(M_\phi z_m) + J_1^2(M_\phi z_m) ]} \frac{\xi (1+G)}{z_m} 
\,. \label{Fphi}
\end{equation}

\subsection{Light techni-dilaton limit} 
\label{sec:LTDlimit}

The physical quantities presented above 
are calculated as functions of three holographic parameters, $\xi, G, z_m$.  
(The UV regulator $\epsilon$ is taken to be zero after the calculations.)   
We shall examine how a light TD can be realized by adjusting these holographic parameters 
and how the presence of the light TD affects other physical quantities.

The light TD limit corresponds to taking $(M_\phi z_m) \ll 1 $ in  Eq.(\ref{MS:FS}) 
such that the eigenvalue equation for the TD mass $M_\phi$ is analytically solved:   
\begin{equation} 
  (M_\phi z_m) \simeq \sqrt{3} \xi  
  \,, 
\end{equation}
which implies $\xi \ll 1$ in the light TD limit. 
 In this limit, the techni-pion decay constant $F_\pi$ in Eq.(\ref{Fpi}) 
 can be approximated as  
\begin{equation} 
  F_\pi \simeq \sqrt{\frac{N_{\rm TC}}{12\pi^2}} \frac{\xi(1+G)}{z_m} 
  \,,  \label{Fpi:approx}
\end{equation} 
so that the TD mass normalized to $(4\pi F_\pi)$ is given as 
\begin{equation} 
  \frac{M_\phi}{4\pi F_\pi} \simeq \sqrt{\frac{3}{N_{\rm TC}}} \frac{\sqrt{3}/2}{1+G} 
  \,. \label{Mphi-Fpi} 
\end{equation} 
This implies 
\begin{equation} 
\frac{M_\phi}{ 4\pi F_\pi} \to 0
\qquad 
{\rm as} 
\qquad 
G \to \infty
\,. 
\end{equation} 
When $M_\phi/(4 \pi F_\pi) \simeq 0.1$, for instance, we find 
\begin{equation} 
  G \simeq (9.9, 8.4, 7.4) 
\,, \qquad 
{\rm for} 
\qquad 
N_{\rm TC} = 3,4,5 
\,.   
\end{equation}

Remarkably, in the light TD limit, 
the TD decay constant $F_\phi$ in Eq.(\ref{Fphi}) normalized to $F_\pi$ in Eq.(\ref{Fpi:approx}) 
becomes completely free from the holographic parameters to be just a constant: 
\begin{eqnarray} 
  \frac{F_\phi}{F_\pi} &\simeq & 
\sqrt{2 N_{\rm TF}} 
\cdot \sqrt{J_0^2(x) + J_1^2(x)} \Bigg|_{x=(M_\phi z_m) \ll 1}   
\nonumber \\ 
&\simeq&   
   \sqrt{2 N_{\rm TF}} 
   \,. \label{Fphi:Fpi}
\end{eqnarray}
 Thus the present holographic model can achieve the limit  
realizing the TD as a 
massless Nambu-Goldstone boson (``conformal limit"):  
\begin{equation} 
  \frac{M_\phi}{4\pi F_\pi} \to 0 
  \qquad 
  {\rm and} 
  \qquad 
  \frac{F_\phi}{F_\pi} 
  \to {\rm finite}
  \,, 
 \qquad  
 {\rm as} 
 \qquad 
 G \to \infty 
 \,. \label{dilaton:limit}
\end{equation}

 In the conformal limit Eq.(\ref{dilaton:limit}) 
the techni-gluon condensate normalized to the fixed $(4 \pi F_\pi)^4$, 
$\langle \alpha G_{\mu\nu}^2 \rangle/( 4 \pi F_\pi)^4$,  goes to infinity (See Eq.(\ref{FFbar})) 
\begin{equation} 
\frac{\langle  \alpha G_{\mu\nu}^2 \rangle}{(4\pi F_\pi)^4} 
\sim G 
\to \infty\,.   
\end{equation}
If the PCDC holds, then the beta function $\beta(\alpha)$ of the TC gauge coupling $\alpha$ in the present holographic 
model would read
\begin{eqnarray}  
\beta(\alpha) 
&=& \frac{\alpha}{\langle G_{\mu\nu}^2 \rangle} F_\phi^2 M_\phi^2  
\nonumber \\ 
 &\sim& 
\frac{1}{G(1+G)^2}  
 \to 0 
 \qquad 
 {\rm as}
 \qquad 
 G \to \infty 
\,. 
\end{eqnarray}
It is interesting to compare these 
result with those of the ladder calculation near the criticality 
$\alpha_* \simeq \alpha_{c}$~\cite{Hashimoto:2010nw}: 
\begin{eqnarray} 
  \frac{  \langle G_{\mu\nu}^2 \rangle}{m_F^4} 
&\sim& 
\frac{\langle G_{\mu\nu}^2 \rangle}{(4 \pi F_\pi)^4} 
\sim 
\left( \alpha_*/\alpha_c - 1\right)^{-3/2} 
\to 
\infty 
\,,  \nonumber\\ 
\beta(\alpha) 
&\sim& 
  \left( \alpha_*/\alpha_c - 1\right)^{+3/2} 
\to 0  
\,, \nonumber \\ 
\frac{\langle \theta_\mu^\mu \rangle}{m_F^4} 
&=& 
\frac{\beta(\alpha )}{\alpha} \times \frac{\langle G_{\mu\nu}^2 \rangle}{m_F^4} 
\to 
{\rm constant} \neq 0 
\qquad 
{\rm as} 
\qquad 
\alpha_* \to \alpha_c 
\,, \label{lad:result}
\end{eqnarray}
where $\alpha_*$ and $\alpha_c$ respectively denote the Caswell-Bank-Zaks infrared fixed point 
of the two-loop beta function for the WTC~\cite{Caswell:1974gg} and 
the critical coupling of the chiral symmetry breaking in 
the ladder approximation. 
 As clearly seen from Eq.(\ref{lad:result}), the divergence of $\langle G_{\mu\nu}^2 \rangle$ 
 precisely cancels with the vanishing $\beta(\alpha)$, so that  
\begin{equation} 
 \frac{F_\phi^2}{m_F^2} \cdot \frac{M_\phi^2}{m_F^2} 
\sim 
\frac{\langle \theta_\mu^\mu \rangle}{m_F^4} 
\to {\rm constant} \neq 0 
\qquad 
{\rm as} 
\qquad 
\alpha_* \to \alpha_c 
\,. 
\end{equation} 
 This results in the no massless limit unless $F_\phi/m_F \to \infty$, i.e., a decoupled TD~\footnote{ 
Incidentally, a parametrically light TD was argued in 
the framework of the ladder approximation~\cite{Bando:1986bg,Dietrich:2005jn,Appelquist:2010gy}:   
It was claimed that 
$F_\phi^2 M_\phi^2/m_F^4 \sim \beta(\alpha) \cdot \langle G_{\mu\nu}^2 \rangle/m_F^4 \to 0$ 
as $\beta(\alpha)$ goes to zero near the criticality, 
based on an assumption that 
$\langle G_{\mu\nu}^2 \rangle/m_F^4 \to $constant $< \infty$, 
which actually contradicts the explicit computation in Eq.(\ref{lad:result}). }.

Given the techni-pion decay constant $F_\pi$ in Eq.(\ref{Fpi:approx}),  
we may express the chiral condensate in Eq.(\ref{FFbar}), 
with the renormalization scale $1/L$ set to $F_\pi$ as 
\begin{equation} 
 \langle \bar{F}F \rangle_{1/L=F_\pi} 
= - c_S N_{\rm TF} \sqrt{\frac{N_{\rm TC}}{3\pi^2}} \frac{F_\pi^2}{z_m} 
\,, \label{FFbar:Fpi:holo}
\end{equation}
with $c_S=\sqrt{3}/2$ in Eq.(\ref{cS}). 
 On the other hand, we may parametrize  $\langle \bar{F}F \rangle_{F_\pi}$ as~\footnote{
 Note that the renormalization scale $\mu=F_\pi$ depends on $N_{\rm TC}$, so that 
$\langle  \bar{F}F \rangle_{F_\pi}$ scales like $\sim N_{\rm TC}^{3/2}$ in a way different from 
$\langle \bar{F}F \rangle_{m_F} \sim N_{\rm TC}$.  
} 
\begin{equation} 
  \langle \bar{F}F \rangle_{F_\pi} 
= - \bar{\kappa}\,  N_{\rm TF} \, 4\pi F_\pi^3 
\,,  \label{FFbar:Fpi:gene}
\end{equation}
where the overall coefficient 
$\bar{\kappa}$ is to be determined once a straightforward nonperturbative calculation is done. 
 From Eqs.(\ref{FFbar:Fpi:gene}) and (\ref{FFbar:Fpi:holo}) we find 
\begin{equation} 
 F_\pi = \frac{\sqrt{N_{\rm TC}}}{8 \pi^2 \bar{\kappa}} \frac{1}{z_m}
 \,.  
\end{equation}
  Comparing this with Eq.(\ref{Fpi:approx}) 
we thus see that the holographic parameters $\xi$ and $G$ are now 
correlated involving $\bar{\kappa}$:  
\begin{equation} 
  \xi (1+G) = \frac{\sqrt{3}}{4\pi \bar{\kappa}}
  \,. \label{xi-G:relation}
\end{equation} 
 For a reference value of $\bar{\kappa}$ in Eq.(\ref{FFbar:Fpi:gene}), 
a recent nonperturbative analysis based on the ladder approximation 
  corresponds to $\bar{\kappa} \simeq 0.16$~\cite{Hashimoto:2010nw}. 
Including this reference value, 
we shall take $\bar{\kappa} = (0.016, 0.16, 1.6)$ such that $\xi$ and $G$ are constrained as 
\begin{equation} 
 \xi (1+G) \simeq (9, 0.9, 0.09)
\,. \label{xi-G:cons}
\end{equation}

\begin{figure}[t] 
\begin{center} 
\includegraphics[width=8.0cm]{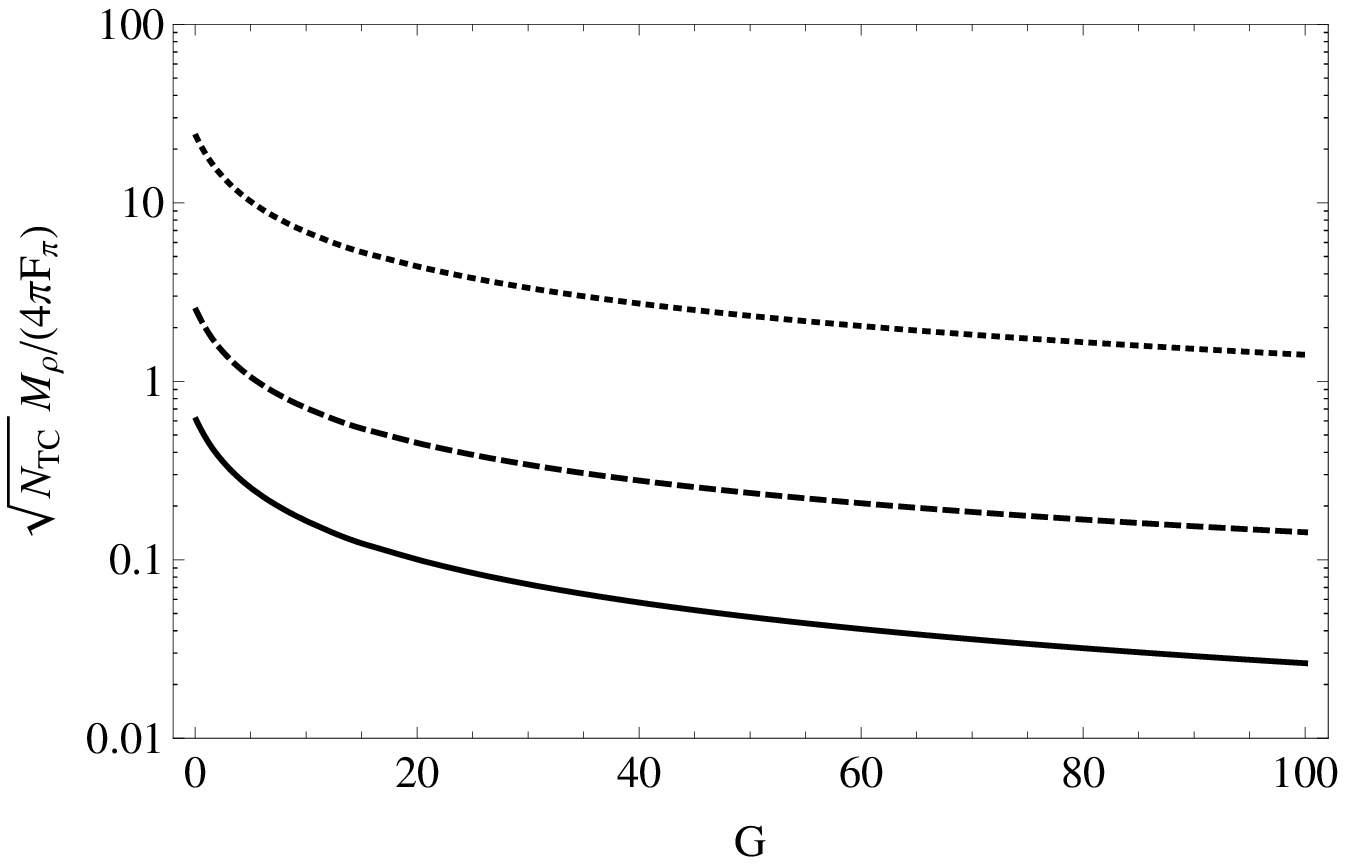}
\includegraphics[width=8.0cm]{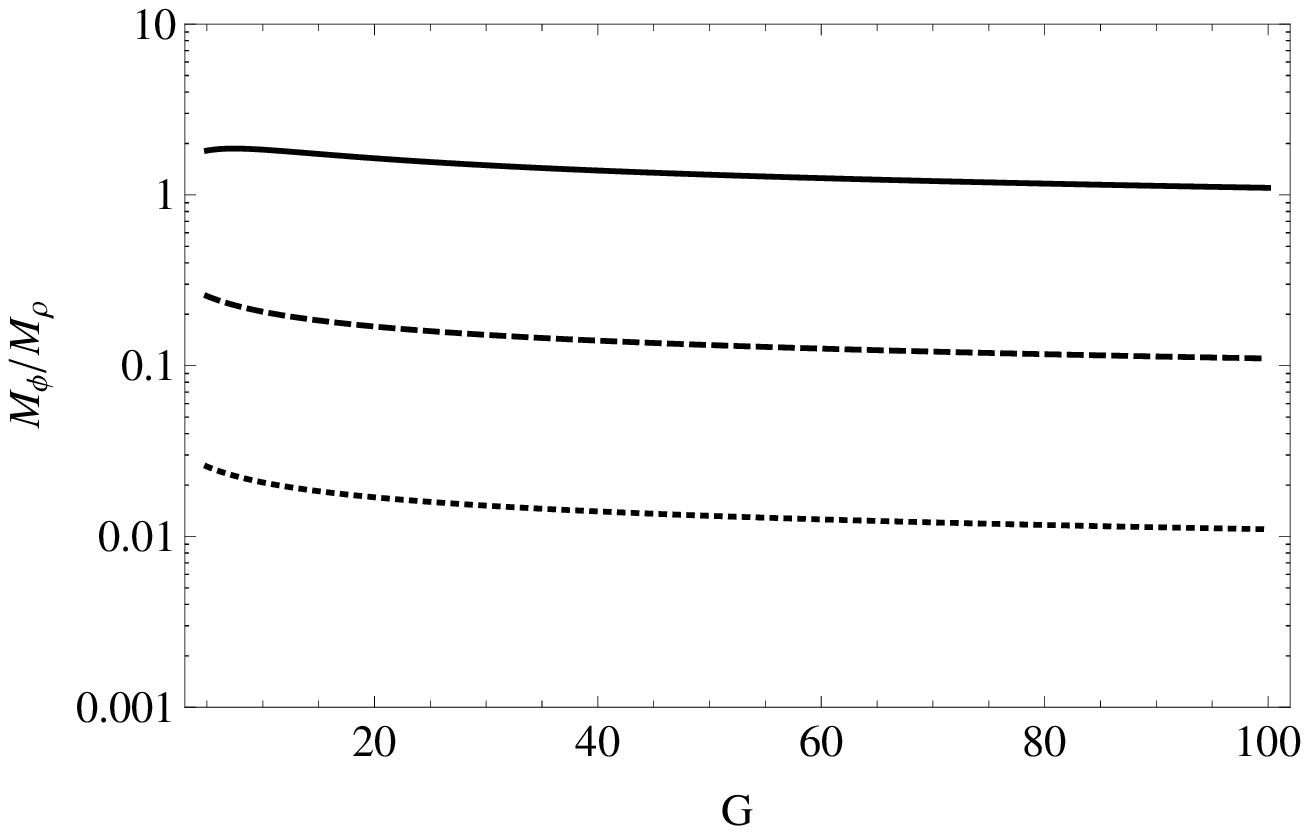}
\end{center}
\caption{ 
Left: The $G$-dependence of $\sqrt{N}_{\rm TC} M_\rho/(4\pi F_\pi)$  
with $\bar{\kappa} =$ 0.016 (solid), 0.16 (dashed) and 1.6 (dotted) fixed. 
Right: The plot of $M_\phi/M_{\rho}$ as a function of $G$ with the same values for $\bar{\kappa}$ taken. 
}  
\label{mass-zm-G:plot}
\end{figure}

 \begin{figure}[t] 
\begin{center} 
\includegraphics[scale=0.6]{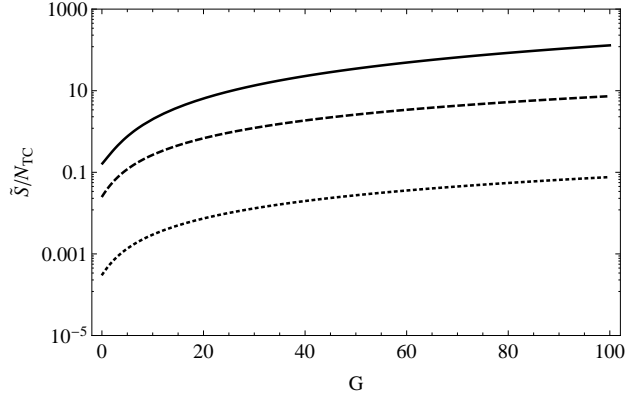}
\end{center}
\caption{ 
The $G$-dependence of $\widehat{S}/N_{\rm TC}$ with $\bar{\kappa} =$ 0.016 (solid), 0.16 (dashed) and 1.6 (dotted)
 fixed. }  
\label{Shat-G}
\end{figure}

We now look into the masses of techni-$\rho$ and -$a_1$ mesons normalized to $(4\pi F_\pi)$,  
$M_\rho/(4 \pi F_\pi)$ and $M_{a_1}/(4\pi F_\pi)$, in the conformal limit Eq.(\ref{dilaton:limit}).  
 Using Eqs.(\ref{Mrho:Ma1}) and (\ref{Fpi:approx}) with Eq.(\ref{xi-G:cons}) 
we see that these ratios can be calculated as a function of the parameter $G$ only:     
\begin{eqnarray} 
 \frac{M_\rho}{4 \pi F_\pi} 
&\simeq& 
\frac{2 \pi \bar{\kappa} \widetilde{M}_\rho(G) }{\sqrt{N_{\rm TC}}} 
\,, \nonumber \\ 
\frac{M_{a_1}}{4 \pi F_\pi} 
&\simeq& 
\frac{2 \pi \bar{\kappa} \widetilde{M}_{a_1}(\xi, G) }{\sqrt{N_{\rm TC}}} \Bigg|_{\xi 
= \frac{\sqrt{3}}{4\pi \bar{\kappa} (1+G)}}
\,. 
\end{eqnarray}
In the conformal limit Eq.(\ref{dilaton:limit}), 
we thus find that $M_\rho/(4 \pi F_\pi) \simeq M_{a_1}/( 4\pi F_\pi)$ 
and goes to zero:  
\begin{equation} 
  \frac{M_\rho}{4\pi F_\pi} \simeq \frac{M_{a_1}}{ 4\pi F_\pi}
\to 0 
  \qquad 
  {\rm as} 
  \qquad 
  G \to \infty 
  \,. 
\end{equation} 
 In Fig.~\ref{mass-zm-G:plot} we plot the $G$-dependence of $\sqrt{N}_{\rm TC} M_\rho/(4\pi F_\pi)$ (left panel). 
The figure shows that the ratio $M_\rho/(4\pi F_\pi)$ slowly gets smaller as $G$ increases and 
finally reaches zero in the conformal limit $G\to \infty$. 
This critical phenomenon looks similar to the vector realization/vector manifestation~\cite{Georgi:1989gp,Harada:2000kb,Harada:2003jx}. 
 Also has been plotted the ratio $M_\phi/M_{\rho}$ as a function of $G$ (right panel). 
Again, the ratios $M_\phi/M_{\rho, a_1}$ slowly become smaller as $G$ increases 
and finally go to zero:  
 \begin{equation} 
   \frac{M_\phi}{M_{\rho,a_1}} \to 0  
   \qquad 
   {\rm as} 
   \qquad 
   G \to \infty 
   \,, 
   \label{divS}
 \end{equation}
   which implies that in such a limit the TD is indeed the lightest particle.

Finally,  we examine the effect on the the $S$ parameter in Eq.(\ref{Spara}) in the conformal limit. 
The $S$ is calculated as a function of the holographic parameters $\xi$ and $G$ 
with the constraint in Eq.(\ref{xi-G:cons}). 
We thus plot the $G$-dependence of $\widehat{S}/N_{\rm TC}=S/(N_D N_{\rm TC})$
in Fig.~\ref{Shat-G}, which implies the large $G$-behavior: 
\begin{equation}
  S \to \infty 
  \qquad 
  {\rm as}
\qquad 
G \to \infty 
\,.  \label{S:limit}
\end{equation}
This scaling can be understood by noting that 
$\widehat{S} \simeq (a/4\pi) \cdot (4 \pi F_\pi/M_\rho)^2 = 4\pi/g^2_{\rm HLS}$~\cite{Haba:2008nz}, 
where $g_{\rm HLS}$ denotes the gauge coupling of the techni-$\rho$ meson regarded 
as a gauge boson of the hidden local symmetry~\cite{Harada:2003jx} 
and $a (\simeq 2)$ is a parameter of the hidden local symmetry model.   
Then $\widehat{S} \to \infty$ limit corresponds to $(4 \pi F_\pi/M_\rho)^2 \to 0$ or 
$g_{\rm HLS} \to 0$ limit (vector realization/vector manifestation)~\cite{Georgi:1989gp,Harada:2000kb,Harada:2003jx}.

\section{Holographic techni-dilaton at 125 GeV}
\label{sec:HTD125}

In this section we discuss the 125 GeV holographic TD  
by matching the present model to the one-family WTC model with $N_D=4$, 
$F_\pi=123$ GeV as a typical example of the WTC.
We write 
\begin{equation}
N_{\rm TF} = 2 N_D + N_{\rm EW-singlet} \ge 2 N_D=8\,,
\end{equation}
where 
$N_{\rm EW-singlet}$ denotes the number of `dummy' techni-fermions 
which are singlet under the electroweak charges 
and only contribute to realizing the walking behavior. 
Actually, most of the variants of the WTC have a tendency similar to the one studied here, 
except for a class of WTC models without colored/charged weak-doublets, e.g.,  
the ``one-doublet model''  ($N_D=1$, $F_\pi=v_{\rm EW}=246$ GeV) 
 which was shown~\cite{Matsuzaki:2011ie,Matsuzaki:2012gd,Matsuzaki:2012vc} to be invisible at LHC due simply to
the smallness of the coupling ($\sim 1/F_\phi \ll 1/v_{\rm EW}$) without compensating enhancement by the colored/charged techni-fermions.

   As seen from Eq.(\ref{Mphi-Fpi}), 
in the one-family model a light TD with the mass around 125 GeV 
is realized when $M_\phi/(4 \pi F_\pi) \simeq 0.1$,  
which corresponds to the parameter $G\simeq 10$:   
\begin{equation} 
M_\phi \simeq 125\, {\rm GeV} 
\qquad 
{\rm at} 
\qquad 
G \simeq 10  
\,. 
\end{equation}
The value of this $G$ is compared to the real-life QCD value $\simeq 0.25$~\cite{Haba:2010hu}. 
As noted in Eq.(\ref{Fphi:Fpi}), in such a light TD case, 
the TD decay constant $F_\phi$ is fixed 
by the techni-pion decay constant $F_\pi$ 
independently of the holographic parameters as well as the number of $N_{\rm TC}$. 
  For $F_\pi = 123$ GeV we thus estimate $F_\phi$ and a ratio $v_{\rm EW}/F_\phi$ to find  
 \begin{eqnarray} 
 F_\phi &\simeq& (514, 630, 730, 813)\,{\rm GeV} 
\,,  \nonumber \\ 
 \frac{v_{\rm EW}}{F_\phi} =\frac{ \frac{N_D}{2}F_{\pi}}{F_\phi} &\simeq&   (0.49, 0.39, 0.33, 0.30) \simeq \sqrt{\frac{2}{N_{\rm NF}}}\quad (N_D=4) \,, 
  \label{Fphi:numbers}
\end{eqnarray} 
for $N_{\rm EW-singlet} = 0, 4, 8, 12$, i.e., $N_{\rm TF}=8, 12, 16, 20$ in accord with Eq.(\ref{Fphi:Fpi}). 
It is remarkable that  the above result is fairly insensitive to a particular value of $M_\phi\simeq 125 {\rm GeV}$, in sharp contrast to the estimate explicitly based on the PCDC \cite{Matsuzaki:2012mk} which 
is very sensitive to $M_\phi$.
Note also that our result Eq.(\ref{Fphi:numbers}) is free from any additional assumption such as the 
ladder criticality condition $N_{\rm TF} \simeq 4 N_{\rm TC}$ used in Ref.\cite{Matsuzaki:2012mk} 
which was based on the ladder approximation.

Once the TD decay constant $F_\phi$ is estimated, 
we are now ready to discuss the LHC phenomenology of the 125 GeV holographic TD in the same way as in Refs.~\cite{Matsuzaki:2012vc,Matsuzaki:2012mk}: 
The TD couplings to the SM gauge bosons are obtained just by 
scaling from the SM Higgs as $v_{\rm EW} \to F_\phi$. 
The coupling to the SM-$f$ fermion, on the other hand, is set by the mass $m_f$ divided by $F_\phi$ 
along with a factor $(3-\gamma_m)$, so that the scaling goes like 
$m_f/v_{\rm EW} \to (3-\gamma_m) m_f/F_\phi$~\cite{Bando:1986bg,Matsuzaki:2012vc,Matsuzaki:2012mk}
. 
The anomalous dimension $\gamma_m$ for the third-generation fermions 
are taken to be $\simeq 2$ so as to realize the realistic fermion masses 
by strong extended TC (ETC) dynamics~\cite{Miransky:1988gk}, 
while we put $\gamma_m \simeq 1$ 
for the other lighter fermions in order to 
avoid excessive flavor changing neutral currents 
(See also Eq.(\ref{mql:2nd})). 
(Throughout the holographic computations described so far, 
we have set $\gamma_m =1$ since the holographic model is thought of as dual to WTC, not involving 
the SM fermion sector concerning a type of ETC.)  We thus have 
\begin{eqnarray} 
  \frac{g_{\phi WW/ZZ}}{g_{ h_{\rm SM} WW/ZZ }} 
  &=& \frac{v_{\rm EW}}{F_\phi} 
  \simeq \sqrt{\frac{2}{N_{\rm TF}}}\,, \nonumber \\
 &\simeq&  \frac{g_{\phi ff}}{g_{h_{\rm SM} ff}}  \,
\quad ({\rm for} \quad f=t,b,\tau) 
\,.  \label{scaling}
\end{eqnarray} 
Thus the processes involving these couplings are suppressed compared with the SM Higgs by the characteristic factor $(v_{\rm EW}/F_\phi)^2 \simeq 2/N_{\rm TF}\ll 1$ for the typical WTC with $N_{\rm TF} \gg 1$. 

On the contrary, 
the couplings to digluon and diphoton 
are largely enhanced compared with the SM Higgs, which somehow compensates the smallness of  other couplings in Eq.(\ref{scaling}) in most of the channels currently studied at LHC as shown before~\cite{Matsuzaki:2012mk}
(see also the discussions in the next paragraph).~\footnote{Note that this kind of enhancement of the $\gamma\gamma$ and $gg$ couplings is generic also for other models having extra heavy fermions such as  the typical fourth generation model 
which however, having the same couplings as that of the SM Higgs, are severely constrained,  in sharp contrast to our case with the suppressed couplings in Eq.(\ref{scaling}).
}  This is  the most characteristic feature of the TD in the generic WTC (having colored/charged techni-fermions) in contrast to other dilaton/radion models as well as
the one-doublet model:
In the case at hand, the one-family model,  these couplings are in fact  
enhanced by the colored/charged 
techni-fermion loop contributions along with a factor $N_{\rm TC}$~\cite{Matsuzaki:2011ie,Matsuzaki:2012vc,Matsuzaki:2012mk}:
\begin{eqnarray} 
 {\cal L}_{\rm eff}^{\gamma\gamma,g g} =\frac{\phi}{F_\phi} \left\{ 
 \frac{\beta_F(g_s)}{2g_s} G_{\mu\nu}^2 + \frac{\beta_F(e)}{2e} F_{\mu\nu}^2 
\right\}\,,\\
 \beta_F(g_s) = \frac{g_s^3}{(4\pi)^2} \frac{4}{3} N_{\rm TC} \,,\quad
\nonumber  
  \beta_F(e) = \frac{e^3}{(4\pi)^2} \frac{16}{9} N_{\rm TC} 
\,.   \label{betas}
\end{eqnarray} 
    We thus find the scaling from the SM Higgs for the couplings to 
$gg$ and $\gamma\gamma$, which can approximately be expressed 
 at around 125 GeV (Detailed formulae are given in the Appendix of Ref.\cite{Matsuzaki:2012vc}):\cite{Matsuzaki:2012mk} 
 \begin{eqnarray} 
\frac{g_{\phi gg}}{g_{h_{\rm SM} gg}} 
&\simeq & 
\frac{v_{\rm EW}}{F_\phi} 
\cdot 
\left( (3-\gamma_m) + 2 N_{\rm TC} \right)   \,,
\nonumber \\ 
\frac{g_{\phi \gamma\gamma}}{g_{h_{\rm SM} \gamma\gamma}} 
&\simeq & 
\frac{v_{\rm EW}}{F_\phi} 
\cdot 
 \left( \frac{63 -  16(3-\gamma_m)}{47} - \frac{32}{47} N_{\rm TC} \right)  
\,,  \label{g-dip-dig}
\end{eqnarray} 
where in estimating the SM contributions  
we have incorporated only the dominant ones, the top (the terms having $3-\gamma_m (=1)$) and the 
$W$ boson (the term of 63/47 for $\gamma\gamma$ rate) loop contributions, which largely cancel each other in the diphoton channel. 
It is thus clear  that the techni-fermion contributions overwhelm those of the SM particles for the $\gamma \gamma$ channel (for $N_{\rm TC} >2$) as well as the $g g$ channel. 

These couplings actually play the key role to account for the presently 
reported excess of diphoton event rate, 
while the significance for other channels stays at the level similar to 
the SM Higgs prediction:  
Although the TD production through the vector boson fusion process is suppressed by an amount of $(v_{\rm EW}/F_\phi)^2$, 
the diphoton rate along with dijet becomes consistent with the current LHC data because of 
the large contamination with the gluon fusion events  which 
are highly enhanced to be about 80$\%$ or more  
in the case of TD compared to the SM Higgs case with $\sim$ 30$\%$, due to the larger gluon fusion cross section. 
 As for other exclusive channels with jets, the current accuracy has not reached a level 
which can more precisely distinguish the production processes than the diphoton channel. 
As the currently most relevant event categories, 
we shall therefore take the $\gamma\gamma 0j$ and $\gamma\gamma 2j$ events in addition to the $b\bar{b}$ channel 
to be exclusive and other channels such as $WW^*, ZZ^*$ and $\tau\tau$ to be inclusive, as was done in Ref.~\cite{Matsuzaki:2012mk}.

We can thus estimate the 125 GeV TD signals at the LHC and perform the goodness-of-fit 
to the currently available data set~\cite{ATLAS-CONF-2012,CMS-PAS-HIG},  
in a way similar to that done in Ref.~\cite{Matsuzaki:2012mk}. 
The best-fit value of $v_{\rm EW}/F_\phi$ found in Ref.~\cite{Matsuzaki:2012mk} is 
$v_{\rm EW}/F_\phi|_{\rm best-fit} \simeq 0.2$ for $N_{\rm TC}=4,5$ as in Eq.(\ref{best-fit}), 
which is slightly off by about 20\%--30\% from the present holographic prediction Eq.(\ref{Fphi:numbers}):
\begin{equation} 
\frac{v_{\rm EW}}{F_\phi} \Bigg|_{\rm holo} \simeq  \sqrt{\frac{2}{N_{\rm TF}}}
\simeq 0.3 -0.5
\, .
\end{equation}  
 However, such $\sim $30\% corrections would come from the next-to-leading order terms in 
$1/N_{\rm TC}$ expansion as was discussed in Ref.~\cite{Harada:2006di}. 
Inclusion of the $1/N_{\rm TC} (\sim  20\% - 30\%)$ corrections for $N_{\rm TC}=3,4,5$   
would then give a shift:  
\begin{equation} 
\frac{v_{\rm EW}}{F_\phi}\Bigg|_{\rm holo} 
\to \frac{v_{\rm EW}}{F_\phi}\Bigg|^{+1/N_{\rm TC}}_{\rm holo} 
\sim 0.2 -0.4 
\, . \label{holo:prediction}
\end{equation}

The holographically predicted $v_{\rm EW}/F_\phi$ in Eq.(\ref{holo:prediction}) 
is also consistent with the ladder estimate~\cite{Matsuzaki:2012mk}, 
$v_{\rm EW}/F_\phi \simeq 0.1-0.3$ in Eq.(\ref{vals})~\footnote{
Also, the predicted numbers in Eq.(\ref{holo:prediction}) roughly coincide 
with the value estimated from other holographic models~\cite{Lawrance:2012cg}. 
}. 
Note that the two calculations are quite 
different qualitatively in a sense that 
the ladder calculation has no massless TD limit, while 
the present holographic model including the nonperturbative 
gluonic dynamics does. 
Nevertheless, such a numerical coincidence 
may suggest that 
both models are reflecting some reality through similar dynamical effects 
for the particular mass region of the 125 GeV TD.

Using the predicted $v_{\rm EW}/F_\phi$ including the possible $1/N_{\rm TC}(=0.3,0.25,0.20)$ corrections 
for $N_{\rm TC}=3,4,5$,  in Table~\ref{tab:chi2} we list the results of the $\chi^2$ fit based on 
the currently available LHC data set~\cite{ATLAS-CONF-2012,CMS-PAS-HIG}. 
The table shows that the current data favors the holographic TD in the one-family model with $N_{\rm TC}=4$ 
and $N_{\rm EW-singlet}=8, 12$ (i.e. $N_{\rm TF}=16, 20$), slightly better than the SM Higgs with $\chi^2/{\rm d.o.f} \simeq 1.0$. 
The upcoming more data will conclude whether the TD is more favorable than the SM Higgs, or not.

\begin{table}[t] 
\begin{tabular}{|c|c|c|} 
\hline 
\hspace{5pt} 
 $N_{\rm TC}$ 
\hspace{5pt} 
& 
\hspace{5pt} $[v_{\rm EW}/F_\phi]^{+1/N_{\rm TC}}_{\rm holo} $ with $N_{\rm EW-singlet}=(0,4,8,12)$ 
\hspace{5pt} 
& 
\hspace{5pt}  
$\chi^2/{\rm d.o.f}$ with d.o.f = 14  
\hspace{5pt} 
\\ 
\hline \hline 
3 & (0.34, 0.27, 0.23, 0.21) & (3.5, 2.1, 2.0, 2.2) \\ 
\hline 
4 & (0.37, 0.29, 0.25, 0.23) & (9.4, 2.1, 1.0, 0.8) \\ 
\hline 
5 & (0.39, 0.31, 0.26, 0.24) & (55, 16, 6.1, 3.7) \\ 
\hline 
\end{tabular} 
\caption{ The results of the $\chi^2$ fit based on 
the currently available LHC data set~\cite{ATLAS-CONF-2012,CMS-PAS-HIG}. 
The data adopted here are the same as those used in the analysis in Ref.~\cite{Matsuzaki:2012mk}. 
The SM Higgs gives $\chi^2/{\rm d.o.f}\simeq 1.0$.  
} 
\label{tab:chi2}
\end{table}

Although it is not relevant to the above analysis of the current LHC data, 
 we may further impose a phenomenological constraint on the $S$ parameter, say, 
\begin{equation} 
S=0.1
\,. 
\end{equation}
Then all the holographic parameters $\xi$ and $z_m$ in addition to $G \simeq 10$ 
are completely fixed to be   
\begin{equation} 
  \xi \simeq 0.014 
  \,, \qquad 
  z_m^{-1} \simeq 5.2 \,{\rm TeV}  
  \,, 
\end{equation} 
for $N_{\rm TC}=4$, 
where we have $\bar{\kappa} \simeq 1.0$ (i.e. $\xi (1+G) \simeq 0.14$ in Eq.(\ref{xi-G:cons})). 
It should be noted that although $S$ is divergent in the conformal limit where TD becomes exactly massless, see Eq.(\ref{divS}), 
$S$ grows extremely slowly  as $G$ increases as can be seen from Fig.\ref{Shat-G},  and hence such a small $S=0.1$ is easily realized for a relatively light TD mass like $\simeq 125$ GeV~\footnote{ 
One might think that such a light TD with the decay constant $F_\phi$ larger than $v_{\rm EW}$ by about 80$\%$ 
is incompatible with the precision electroweak test like $S$ and $T$ parameters~\cite{Barbieri:2007bh}.  
However, such an argument is restricted to the low energy effective theory, because ultraviolet contributions coming from 
heavier mesons like techni-$\rho$ would compensate the TD contribution to be consistent with the $ST$ bounds, 
as was pointed out in Ref.~\cite{Campbell:2011iw} (See also a comment in a paper~\cite{Bellazzini:2012vz} which appeared after submission of our paper). 
Actually, our calculation includes a full non-perturbative TC dynamics not just TD and hence is  a concrete example to realize such a compensation. 
}.
The estimated numbers of the 
holographic parameters for matching to the one-family models with $N_{\rm TC}=3,4,5$ are summarized 
in Table~\ref{tab:numbers:1}.

\begin{table}[t] 
\begin{tabular}{|c||c|c|c||c|c|c|c|} 
\hline 
$N_{\rm TC}$  & 
$M_\phi$ [GeV] (input) & 
$F_\pi$ [GeV] (input) & 
$S$ (input) & 
\hspace{5pt} $G$ \hspace{5pt} & 
\hspace{5pt}  $\xi$ \hspace{5pt} & 
\hspace{5pt}  $z_m^{-1}$ [TeV] \hspace{5pt}  & 
\hspace{5pt}  $\bar{\kappa} = \frac{\sqrt{3}}{4 \pi \xi(1+G)}$ \hspace{5pt}  \\ 
\hline \hline 
3 & 125 & 123 & 0.1 & 10 & 0.014 & 5.2 & 0.89 \\ 
4 & 125 & 123 & 0.1 & 8.7 & 0.015 & 4.8 & 0.96 \\ 
5 & 125 & 123 & 0.1 & 7.7 & 0.016 & 4.5 & 0.96 \\ 
\hline 
\end{tabular}
\caption{The holographic parameters estimated by fixing $M_\phi=125$ GeV, $F_\pi=123$ GeV and $S=0.1$ 
for the one-family WTC with $N_{\rm TC}=3, 4,5$. }
\label{tab:numbers:1}
\end{table}

Implications of this parameter-set can be seen in 
a typical mass of the SM fermion (the second-generation lepton and quarks): 
The SM fermion masses are generated through an ETC  
induced four-fermion interaction  to be $m_f \sim - \langle \bar{F} F \rangle_{\Lambda_{\rm ETC}}/\Lambda_{\rm ETC}^2 \sim 
- (\Lambda_{\rm ETC}/F_\pi) \langle \bar{F} F \rangle_{F_\pi}/\Lambda_{\rm ETC}^2$, where $\Lambda_{\rm ETC}$ is the ETC scale 
taken to be $\gtrsim (10^3-10^4)$ TeV to avoid excessive flavor-changing neutral currents among the  second-generation SM fermions. 
 Using Eq.(\ref{FFbar:Fpi:gene}) with $\bar{\kappa} \simeq 1.0$ and $F_\pi=123$ GeV 
one can obtain 
\begin{eqnarray} 
m_{q,l} 
&\sim&  
\bar{\kappa} \cdot N_{\rm TF} \frac{4 \pi F_\pi^2}{\Lambda_{\rm ETC}} 
\nonumber \\ 
&\sim& 
100\, {\rm MeV} - 
1 \, {\rm GeV}
\,,   \label{mql:2nd}
\end{eqnarray}   
 for $N_{\rm TF}=8-20$ and $\Lambda_{\rm ETC} = 10^3-10^5$ TeV.

Taking the parameter-set listed in Table~\ref{tab:numbers:1},  
we completely estimate other physical quantities presented in the previous section.  
In Table~\ref{tab:numbers:2} we list the numbers for 
other physical quantities estimated by taking $S=0.1$, $F_\pi=123$ GeV and $M_\phi=125$ GeV.   
Table~\ref{tab:numbers:2} shows that the TD is indeed the lightest particle, 
 lighter than other TC hadrons such as techni-$\rho$, $-a_1$ and -glueball which 
are on the order of TeV scale, as was discussed in Ref.~\cite{Haba:2010hu}. 
 The dynamical mass of techni-fermion $m_F$ has been estimated in the following way: 
The scale $m_F$ may be defined through the chiral condensate renormalized at $\mu=m_F$: 
$
\langle \bar{F}F \rangle_{m_F} = - \kappa_c \cdot \frac{ N_{\rm TF}N_{\rm TC}}{4\pi^2} m_F^3     
$, 
with the overall coefficient $\kappa_c$ similar to $\bar{\kappa}$ in Eq.(\ref{FFbar:Fpi:gene}). 
One can scale this up to $\mu=F_\pi$ by the scaling law of the chiral condensate with 
the anomalous dimension $\gamma_m=1$, 
$ 
\langle \bar{F}F \rangle_{m_F} = (m_F/F_\pi) \langle \bar{F}F \rangle_{F_\pi} 
$.  
Using Eq.(\ref{FFbar:Fpi:gene}) one thus gets  
$m_F = \sqrt{\pi\bar{\kappa}/\kappa_c} (4 \pi F_\pi/\sqrt{N_{\rm TC}})$ to estimate $m_F \sim 1$ TeV with 
a reference value $\kappa_c=2.0$ used~\footnote{
Numerically, $\kappa_c$ coincides with $(3-\gamma_m)$ 
for $\gamma_m \simeq 0,1,2$ when a simple-minded ansatz for the mass function of techni-fermion $\Sigma(-p^2)$, 
$\Sigma(p^2) \approx m_F (p^2/m_F^2)^{(\gamma_m-2)/2}$ for $p^2 > m_F^2$ and $\Sigma(p^2)=m_F$ for $p^2 < m_F$ 
in evaluating the chiral condensate~\cite{Haba:2010hu}. 
In the case of QCD with $\gamma_m \simeq 0$, 
this ansatz implies the dynamical quark mass $m_q \simeq 453$ MeV 
for the value of $\langle \bar{q}q \rangle \simeq - (277\,{\rm MeV})^3$ 
which is in accord with the conventional constituent quark mass $m_q \simeq 350$ MeV. 
\label{footnote:simple-minded}
}. 
The situation with such a $m_F \sim 1$ TeV suits well with working on 
the effective TD nonlinear Lagrangian formulated in Ref.~\cite{Matsuzaki:2012vc}
.     
 Note also that the masses of techni-$\rho$ and -$a_1$ are slightly different from the simple-minded 
size $M_{\rho, a_1} \sim 2 m_F$, which is due to the presence of the techni-gluon contribution parametrized by 
the holographic parameter $G$. 
 The moderately large $M_{\rho, a_1}$ such as listed in Table~\ref{tab:numbers:2} 
yield a relatively small $S$ parameter even in the light TD case, in contrast to the case 
extremely close to the conformal limit where $S \sim (4 \pi F_\pi)^2/M_\rho^2 \to \infty$ (See Eq.(\ref{S:limit})).

 If we chose $S=0.01$ instead of $S=0.1$, 
we would get $M_\rho \simeq M_{a_1} \simeq 9.8$ TeV, 
 $M_G \simeq 54$ TeV, $F_G \simeq 393$ TeV and $m_F \simeq 1.8$ TeV for $N_{\rm TC}=3$. 
  The large sensitivity for $M_{\rho, a_1, G}$ and $F_G$ comes from 
  the high dependence of $z_m$ on $\xi$, which gets a large shift 
from $S=0.1$ to $S=0.01$ by a factor of about 2.5: 
$z_m^{-1} \simeq 5.2$ TeV $\to$ $z_m^{-1} \simeq 14$ TeV according to a shift 
by a factor of about 1/3 for $\xi$: $\xi \simeq 0.014 \to \xi \simeq 0.005$.  
On the other hand, the parameter $G$ is fairly stable against $S$ to keep $G \simeq 10$ 
because it is almost completely determined by the lightness of the TD (See Eq.(\ref{Mphi-Fpi})). 
 Then $m_F$ gets larger by a factor of about $\sqrt{3}$ simply because 
$m_F \propto \sqrt{\bar{\kappa}} \propto 1/\sqrt{\xi}$.  
Note again that the prediction to $F_\phi$ in Eq.(\ref{Fphi:numbers}) 
is intact whatever smaller value of $S$ we could choose, though 
the predicted numbers for other quantities as above will be somewhat sensitive to the change.

\begin{table}[t]
\begin{tabular}{|c||c|c|c||c|c|c|c|c|}
\hline 
$N_{\rm TC}$  & 
$M_\phi$ [GeV] (input) & 
$F_\pi$ [GeV] (input) & 
$S$ (input) & 
$M_\rho$ [TeV] & 
$M_{a_1}$ [TeV]  & 
$M_G$ [TeV]  &
$F_G$ [TeV]  & 
$m_F$ [TeV] 
\\ 
\hline \hline   
3 & 125 & 123 & 0.1 & 3.5  & 3.5 & 19 & 135  & 1.0 \\   
4 & 125 & 123  & 0.1  & 3.6 & 3.6 & 18   & 156 & 0.97 \\
5 & 125 & 123  & 0.1  & 3.6 & 3.6 & 17   & 174 & 0.87 \\
\hline 
\end{tabular}
\caption{ Other predictions obtained by making a phenomenological input 
for the $S$ parameter $S=0.1$, in addition to setting $F_\pi = 123$ GeV, 
the TD mass $M_\phi=125$ GeV. 
In estimating $m_F$ we have put $\kappa_c =2$ for a reference value (See footnote~\ref{footnote:simple-minded}). } 
\label{tab:numbers:2}
\end{table}

\section{Summary} 
\label{sec:summary}

In summary, 
we reanalyzed a holographic WTC model proposed in Ref.~\cite{Haba:2010hu} 
which incorporates the fully nonperturbative gluonic dynamics, 
in contrast to the ladder approximation.    
Thanks to the full inclusion of the gluonic dynamics,  
we found a limit (``conformal limit"),   
where the TD  becomes a 
massless Nambu-Goldstone boson 
for the scale symmetry spontaneously broken with nonzero and finite TD decay constant $F_\phi$, 
which is never realized in the ladder approximation.

In such a light TD case,  furthermore, 
we found a novel relation  
between the TD decay constant $F_\phi$ and $F_\pi$ (Eq.(\ref{Fphi:Fpi})) 
independently of holographic parameters, 
which unambiguously determines the TD couplings to the SM particles set by 
$v_{\rm EW}/F_\phi$. 
Note that our result  is free from any additional assumption such as the 
ladder criticality condition $N_{\rm TF} \simeq 4 N_{\rm TC}$ used in Ref.\cite{Matsuzaki:2012mk} 
which was based on the ladder approximation.

We then discussed the 125 GeV holographic TD at the LHC taking the one-family model as a definite benchmark. 
The TD couplings to the SM particles set by the ratio $v_{\rm EW}/F_\phi$ were 
estimated, say, for $N_{\rm TC}=4$ and $N_{\rm TF} = 8 + N_{\rm EW-singlet}=16, 20$,  
to be $v_{\rm EW}/F_\phi \simeq 0.2$ (up to $1/N_{\rm TC}$ corrections), 
which turned out to be on the best-fit value in Eq.(\ref{best-fit}) favored by the current data on 
a new boson at 125 GeV recently observed at the LHC~\cite{ATLAS-CONF-2012,CMS-PAS-HIG} (See Table~\ref{tab:chi2}). 
It was shown that the holographically predicted $v_{\rm EW}/F_\phi$ 
in Eq.(\ref{holo:prediction}) 
is also consistent with the ladder estimate 
$v_{\rm EW}/F_\phi \simeq 0.1-0.3$ in Eq.(\ref{vals}). 
Although the two calculations are quite 
different qualitatively in a sense that 
the ladder calculation has no massless TD limit, 
such a numerical coincidence 
may suggest that 
both models are reflecting some reality through similar dynamical effects 
for the particular mass region of the 125 GeV TD.

We further fixed 
all the three holographic parameters by an extra input for the $S$ parameter $S=0.1$. 
Then the present holographic model predicted the masses of the techni-$\rho$, -$a_1$ and techni-glueball 
$M_\rho \simeq M_{a_1} \simeq 3.6$ TeV, $M_G \simeq 18$ TeV and 
the techni-glueball decay constant $F_G \simeq 156$ TeV, 
and the dynamical mass of techni-fermion $m_F \simeq 1.0$ TeV $(\simeq 4\pi F_\pi)$, 
for $N_{\rm TC}=4$ (See Table~\ref{tab:numbers:1}).

 Finally, we shall make some comments on the ``conformal limit": 
In the previous work~\cite{Haba:2010hu}, actually, 
it was addressed that there is no massless-dilaton limit for the TD, in contrast to the present result in Eq.(\ref{dilaton:limit}).  
The previous conclusion was deduced from  
assuming the PCDC in the ladder approximation, by which  
the TD decay constant was calculated through the PCDC relation as in Eq.(\ref{PCDC:1}) to be 
$F_\phi^2 = 3 \eta m_F^4/M_\phi^2$ as a function of the TD mass $M_\phi$. 
 In the present work, on the other hand, 
 the $F_\phi$ was computed directly through its definition tied with the spontaneously broken dilatation current 
and the scalar current correlator related to $F_\phi$ by the Ward-Takahashi identities (See Eq.(\ref{Fphi:formula})). 
The result in Eq.(\ref{dilaton:limit}) is therefore a more generic and purely holographic prediction 
without invoking any approximations like the ladder approximation as in the previous study. 
  In the present study, however, the PCDC relation has not explicitly been checked  simply 
because there is no source for the trace of energy-momentum tensor $\theta_\mu^\mu$ in the 
present model, which is a problem to be studied in the future.

\section*{Acknowledgments}

We would like to thank M.~Hashimoto for collaboration at the early stage of this work.  
S.M. is grateful to D~.K.~Hong and M.~Piai for fruitful discussions during his stay at 
Pohang, Korea, for the APCTP focus program.  
This work was supported by 
the JSPS Grant-in-Aid for Scientific Research (S) \#22224003 and (C) \#23540300 (K.Y.).

\end{document}